# Giant two-phonon Raman scattering from nanoscale NbC precipitates in Nb

C. Cao[1,2], R. Tao[3], D. C. Ford[2], R. F. Klie[3], T. Proslier[2], L. D. Cooley[4], A. Dzyuba[4], P. Zapol[2], M. Warren[1], H. Lind[5] and J. F. Zasadzinski[1,2,] [†]

[1] *Physics Department, Illinois Institute of Technology, Chicago, Illinois 60616, USA*

[2] *Materials Science Division, Argonne National Laboratory, Argonne, Illinois 60439, USA*

[3] *Physics Department, University of Illinois at Chicago, Chicago, Illinois, 60607, USA*

[4] *Superconducting Materials Department, Technical Division, Fermi National Accelerator Laboratory, Batavia, Illinois, 60510 USA*

[5] *Department of Physics, Chemistry and Biology (IFM), Linköping University, SE-581 83 Linköping, Sweden*

March 9, 2015

**ABSTRACT**

High purity niobium (Nb), subjected to the processing methods used in the fabrication of superconducting RF cavities, displays micron-sized surface patches containing excess carbon. High-resolution transmission electron microscopy and electron energy-loss spectroscopy measurements are presented which reveal the presence of nanoscale NbC coherent precipitates in such regions. Raman backscatter spectroscopy on similar surface regions exhibit spectra consistent with the literature results on bulk NbC but with significantly enhanced two-phonon scattering. The unprecedented strength and sharpness of the two-phonon signal has prompted a theoretical analysis, using density functional theory (DFT), of phonon modes in NbC for two different interface models of the coherent precipitate. One model leads to overall compressive strain and a comparison to *ab-initio* calculations of phonon dispersion curves under uniform compression of the NbC shows that the measured two-phonon peaks are linked directly to phonon anomalies arising from strong electron-phonon interaction. Another model of the extended interface between Nb and NbC, studied by DFT, gives insight into the frequency shifts of the acoustic and optical mode density of states measured by first order Raman. The exact origin of the stronger two-phonon response is not known at present but it suggests the possibility of enhanced electron-phonon coupling in transition metal carbides under strain found either in the bulk NbC inclusions or at their interfaces with Nb metal. Preliminary tunneling studies using a point contact method show some energy gaps larger than expected for bulk NbC.

† Corresponding author  zasadzinski@iit.edu

# I. INTRODUCTION

Superconducting transition metal nitrides and carbides have been intensely studied due to their physical properties of high hardness and chemical stability as well as transition temperatures, $T_c$, above that of Nb [1]. Potential devices utilizing thin films of transition metal compounds, which might replace those currently using Nb, continue to draw attention [2-5]. Some transition metal carbides (e.g. NbC, TaC) are of particular interest as neutron scattering has revealed several pronounced anomalies (dips) in the phonon dispersions, at particular *q* vectors, that are due to the renormalization effects of strong electron-phonon interaction [6], usually linked to nesting features of the Fermi surface [7,8]. Such compounds serve as a natural testing ground for *ab initio* calculations of the electronic band structure, lattice dynamics, electron-phonon (e-ph) interaction and superconducting $T_c$ and thus remain an active area of investigation [9-11]. New theoretical methods incorporating correlation effects are also being applied, leading to renewed interest in classic, e-ph mediated superconductivity, especially for tailored, functional materials [12].

During the early 1980's it was discovered that Raman spectroscopy was providing important insights into the e-ph interaction and superconductivity. Two-phonon Raman scattering, a higher-order and therefore generally weaker process, was found to be strong in transition metal compounds with relatively high $T_c$ (e.g. NbC, TaC) but absent in low $T_c$ materials, (e.g. HfC) [13]. Subsequent theoretical investigations of the Raman matrix element showed that peaks in two-phonon Raman should originate in the same electronic scattering processes that give rise to phonon renormalization effects [14] and are related to the electron-phonon spectral function, $\alpha^2F(\omega)$, measured by tunneling [15,16] and responsible for superconductivity. A correlation between two-phonon Raman and phonon anomalies has also been suggested for NbN [17].

Here we show that the two-phonon Raman scattering from NbC can be unusually strong and sharp, much stronger than previously observed on bulk samples. The Raman spectra have been observed in micron-sized surface patches on processed Nb that have previously been identified by scanning electron microscopy (SEM) as containing excess carbon to a depth of at least 1 μm. [18] Samples studied include cut-outs from hot spot locations of tested Nb superconducting radiofrequency (SRF) cavities [18] as well as Nb rods and foils subjected to one or more of the processing steps used in SRF cavity construction. The Raman spectra display a very high degree of reproducibility across the various samples studied suggesting that the mechanism of NbC formation is easily reproduced. The carbon appears to originate from the bulk of the Nb, and migrates to the surface, however the exact conditions required for NbC formation are not understood.

The NbC revealed in high resolution scanning transmission electron microscopy (STEM) is in the form of nanoscale inclusions near the surface of strained, chemically-polished, and annealed Nb, that are coherent with the Nb host lattice. The coherency might be linked to the unprecedented strength and sharpness of the two-phonon signal in the NbC. Therefore we present two possible orientations of the fcc NbC structure (one compressive, one tensile) that would allow coherency with bcc Nb and examine these models further using density functional theory (DFT). Assuming that the NbC precipitates are under uniform compressive strain, it can be shown directly that the two-phonon peaks correspond to the regions of phonon anomalies. This model also accurately accounts for the measured upward shift of the acoustic mode density of states (DOS) measured by the defect-induced first order Raman, but cannot explain the absence of a similar strength shift for the optical modes. Thus, we also present a separate computational model of the extended interface between Nb and NbC that accounts for a close lattice match between the two materials and leads to tensile strain in the NbC. This second model provides a phonon DOS at the interface, which is compatible with the first order Raman response and suggests the two-phonon Raman may originate in the localized vibrational modes at such an interface. Finally, we present preliminary point contact tunneling spectroscopy data from these NbC regions where superconducting gaps are found that are larger than those found on the pure Nb regions and in some cases larger than expected for bulk NbC.

**II. IDENTIFICATION OF NIOBIUM CARBIDE INCLUSIONS BY TEM AND EELS**

The specimens investigated here were discovered by serendipity. NbC is an occasional by-product of preparing superconducting radio-frequency (SRF) cavities from high-purity Nb metal by an empirical recipe that includes deep-drawing (~ 50% total mechanical strain), acid polishing, and annealing. Materials science investigations of either cut-outs of tested SRF cavities at failure locations (hot spots) [18,19], or from pristine Nb plates, foils, etc. prepared like SRF cavities [20], have found that such processing of Nb can lead to micron-sized surface patches, easily seen as dark spots under an optical microscope, that contain high amounts of carbon as revealed by Raman and SEM [18]. While the mechanism of NbC formation is still under investigation, combinations of strain and annealing reproducibly produced carbon and NbC clusters, more readily so in samples that received chemical polish. The affinity of carbon for linear defects [21,22] might be important for this process. The present study is novel in that high-resolution TEM and electron energy loss spectroscopy (EELS) have been coupled to study these excess C patches.

The Nb host material used for TEM is described in [20]. The particular specimen came from a high purity, cavity-grade (< 20 ppm carbon by mass) rectangular rod (2 × 2 × 75 mm$^3$) cut by discharge machining from a polycrystalline plate (grain size >45 $\mu$m) in a recrystallized state [23], and strained to 56% total elongation by a tensile stress test machine. The ends of the rod extended past the grip points, and presumably remained unaffected by tensile strain. However, the rod was not tapered and thus it is possible that strain fields existed under the grip points. The entire rod was electropolished to remove 180 μm of metal as in [24], ultrasonically cleaned with a hydrocarbon-removing detergent, rinsed in ultra-pure water, and subsequently annealed at 800 °C in high vacuum for 2 hours. Such processing re-traces the steps used in SRF cavity construction [20]. Two pieces of the rod, one from the middle strained region, sample #1, and one from the end (minimal strain), sample #2, were studied. These samples were scanned by SEM, and surface patches that showed high C content were cut out by focused Ga-ion beam (FEI Helios Nanolab 600) for TEM analysis. These TEM/EELS analyses were conducted in an aberration-corrected cold-field emission instrument (JEOL ARM200CF) at the University of Illinois at Chicago.

Figure 1(a) shows an aberration-corrected STEM annular dark field (ADF) image from the surface of sample #2. A thin surface layer ~ 90 nm thick [as indicated by the broken yellow line in Figure 1(a) inset] is observed which displays a clear difference in contrast. However, the interface between the surface layer and the bulk Nb appears to be diffuse rather than atomically abrupt. In addition, there is no sign of an amorphous surface oxide layer that is often found in pure Nb samples. Such Nb oxide layers are easily detected by TEM contrast images and EELS. Both low-loss (Fig. 1(b)) and core-loss (Fig. 1 (c)) EELS show different spectra in the two regions. In particular, a low energy feature near 10 eV, characteristic of bulk Nb, is absent in the surface layer. The carbon content in the surface region is approximately 50% and EEL spectra indicate it is ordered NbC. Figure 1(a) also shows an atomic-resolution ADF image (Nb atoms only) of the boundary between the two different regions as indicated by the broken line. The bulk Nb is shown in the [111] projection and the interatomic spacing along the (110) direction is measured from the image to be (2.3±0.1) Å, in good agreement with the expected unit-cell dimensions for bcc Nb. While the interface between the Nb and NbC is not abrupt, it has to be pointed out here that there is no apparent change in lattice parameter crossing the boundary. The

Fourier transforms (FFT) of the Nb and NbC regions are shown in Figure 1(d), where the FFT of Nb is shown in green and the NbC in red. The overlap is indicated by yellow. It can be clearly seen from this image that the crystal symmetry and the unit-cell dimensions are very nearly identical in the Nb and NbC regions. Thus, the fcc NbC is a coherent precipitate, commensurate with the bcc Nb matrix.

In Fig. 2(a) is shown an annular bright field (ABF) image from the surface of sample #1. Three inclusions are observed with characteristic dimensions from ~60 nm (middle) to ~1μm (top). An expanded view of the middle inclusion is shown in Fig. 2(b) showing a clear image contrast. In contrast to the data shown in Figure 1, the inclusion appears to be atomically sharp and a clear boundary between the bulk Nb and the lighter inclusion is visible. A core-loss EEL spectrum image is taken from the region indicated by the green box in Fig. 2(b) and the corresponding integrated intensities for the Nb M-edge, the C K-edge and the O K-edge are also shown. Figure 2(c) shows an atomic-resolution ADF image of the interface between the inclusion and the Nb [111] bulk, again with no visible change in crystal structure crossing the boundary. The FFTs from the two different regions are shown in Figure 2(f) again demonstrate that the lattice parameter and crystal symmetry do not change significantly across the interface. However, it appears that the lattice parameter in the NbC is slightly larger compared to that measured in the Nb region. In Figure 2(g) is shown an image intensity profile across the hetero-interface, perpendicular to the Nb(110) planes. Firstly, the decrease in image intensity from Nb to NbC can be clearly seen, since the intensity scales directly with the average atomic number of the material imaged. Secondly, the interatomic spacing in the Nb region is measured to be (2.3±0.1) Å, while the spacings in the NbC have increased to (2.5±0.1) Å. However, the lattice appears to be continuous across the interface with no visible dislocations. The EEL spectra from these two regions are shown in Fig. 2 (d) and 2(e). Here, the red curve is characteristic of pure Nb whereas the black curve shows peaks characteristic of NbC including the sharp C peak near 288 eV.

The conclusion from the STEM studies is that NbC may have resulted from residues of sheet processing, electropolishing and annealing, with distinct precipitates appearing in regions of Nb with added tensile strain. Such observations are consistent with studies of NbZrC alloys with low concentrations of Zr (~1%) and C (~0.4%), which show that with annealing only, NbC precipitates form throughout the Nb matrix that are very stable [25]. However, the results here show that the NbC near the surface is coherent with the host Nb. Friction with rollers, and injection of hydrogen during electropolishing, are known sources of strain at the niobium surface that are not present in the bulk. The observations of coherent NbC precipitates in the two different Nb rods suggests that this type of NbC inclusion is common.

### III. RAMAN SPECTRA OF NIOBIUM CARBIDE INCLUSIONS

Raman spectroscopy was performed on many types of as-processed Nb including cavity cut-outs, strained/annealed foils, and bars similar to samples #1, #2 discussed above. All Raman measurements were done using a Renishaw, inVia, Raman microscope with an unpolarized 785 nm laser source, 30 s exposure time, 100% laser power [27 mW], and a 50X objective. The laser spot size is ~3 μm diameter, ideal for examining these rough surface patches which have lateral dimensions from 10-100 μm. Previous Raman studies of these patches revealed spectral peaks identified as amorphous C and chain-type hydrocarbons [18]. As will be shown, the hot spot cut-outs studied in ref. 18, also revealed NbC Raman spectra.

In Fig. 3(c) are shown a set of Raman spectra from regions of a processed Nb sample that exhibited a number of surface patches, examples of which are shown in the photographs in Fig. 3(a). This particular Nb sample A was processed at Jefferson Laboratory (JLab) according to the following procedures: spark cut from a large grained Nb plate, buffered chemical polish, electropolish (EP), vacuum anneal at 600°C for 10 hours, and a final EP. There appear to be three main peaks in Fig. 3(c) near 12 THz although the highest frequency peak is a doublet indicating two modes close in frequency and not separately resolved. These peaks are highly reproducible and are observed in a variety of patches on this sample. Other processed Nb samples showed similar spectra. Examples are shown in the middle panel of Fig. 3(c) from a "hotspot" cut-out of a Nb cavity [18] that exhibited a medium field Q drop (see [26] for details). In general, these various Nb samples displayed relatively broad additional Raman peaks due to amorphous carbon only [18]. These peaks are also seen in the spectra presented in Fig. 3(c) but at higher wavenumber and not shown. The presence of carbon in the Raman spectra suggested the sharper peaks might be due to an ordered Nb carbide phase and this is confirmed by comparison to published Raman spectra on bulk NbC [13] as shown in Fig. 3(b) where a particular spectrum from this work is presented for comparison.

The band of peaks near 6 THz are similar in location and shape to the acoustic phonon DOS observed obtained from neutron scattering [27] and observed in earlier Raman studies of bulk NbC [13,27]. However, the Raman band here is shifted to higher frequencies, and considering previous theoretical investigations [11], this suggests a possible volume compression of our precipitates. The optical phonons centered near 18 THz also appear shifted to higher frequencies compared to the neutron DOS, but not as much, proportionately, as the acoustic DOS. It should be noted that first order Raman is normally forbidden in NbC and thus the observation of the full phonon DOS is indicative of disorder-induced first order Raman suggesting C vacancies are responsible, as found in bulk NbC [13].

The sharp peaks of the present study near 12 THz show good agreement with a similar band in bulk NbC which has been identified as two-phonon Raman arising from acoustic plus acoustic, (A+A) processes (shaded region) [13]. In particular, a closer examination of the A+A region in bulk NbC reveals 3 subtle peaks, which agree quite well with the peaks observed here. But whereas this band is a weak effect in bulk NbC it is by far the dominant feature of the present study, strongly exceeding the first order Raman signal. This is not unphysical as the first order response should be zero by symmetry and exists only due to disorder [13]. Examination of all of the spectra of Fig. 3(c), for the two differently prepared Nb samples, shows that this anomalously strong two-phonon band is a characteristic feature of these NbC precipitates. Furthermore, the sharpness of the A+A peaks consistently reveal that highest frequency peak is really a doublet indicating two distinct peaks. The highest band of modes near 24 THz is also in good agreement with bulk NbC and has been identified as acoustic plus optical, A+O two-phonon channel [13]. For this band there does not appear to be the strong enhancement but nevertheless we observe three distinct peaks, whereas the bulk NbC data [13] shows a broad band without much structure.

Summarizing the experiments, we have performed Raman and STEM measurements on surface patches (e.g. Fig. 3(a)) on a variety of processed Nb samples. While all such patches reveal excess carbon in SEM and Raman, some of them display Raman spectra consistent with previous studies of bulk NbC but with a strongly enhanced two-phonon signal. Since the STEM studies reveal NbC as a coherent precipitate in the two particular Nb samples studied, it is tempting to link the anomalously strong two-phonon Raman signal to this unique, commensurate structure. Therefore, in the next section we consider in more detail some possible atomic structure models for the NbC precipitate along with DFT calculations of phonon DOS and phonon dispersions for comparison to the Raman spectra.

## IV. FIRST PRINCIPLES CALCULATIONS OF PHONON DISPERSIONS AND DOS

We consider two possible models for a commensurate interface between the niobium carbide precipitates and the host niobium. In the first model, the (111) plane of the NbC is parallel to the (110) plane of Nb. The atomic arrangements and interatomic distances for the Nb atoms in these planes based on bulk lattice parameters are shown in Fig. 1(e). Considering a five-atom, body centered rectangle in the (111) plane in NbC, this can be made commensurate with a similar rectangle in (110) Nb by an elongation of the short side from 3.16 Å to 3.30 Å and a compression of the long side from 5.47 Å to 4.67 Å leading to an overall area compression ratio of 0.89. For the volume, if we force the d-spacing of the (111) planes of NbC to match that of the (110) planes of Nb there is a further contraction leading to an overall volume compression ratio of 0.81. But this assumes the intrinsic lattice constant for the NbC is 4.47 Å, which is the bulk, stoichiometric value. While we do not have independent, exact measurements of the carbon concentration, there are likely C vacancies as evidenced by the first-order Raman spectra, and such defects will reduce the lattice constant. Some NbC inclusions in steels [28] have been found with lattice constant $a$ = 4.36 Å, and using this value increases the volume compression ratio of our precipitates to 0.87. Since the overall strain on the NbC precipitate in this model is quite large and compressive, the focus of our analysis will be on the effect of a large, *uniform* compressive strain on the phonon dispersions and DOS of NbC.

The second possibility considered is that the (110) plane of NbC is parallel to the (110) plane of Nb. This arrangement is commensurate under a relatively small tensile strain in NbC of 1.04, which is in-plane isotropic, resulting in elongation of the short side from 3.16 to 3.30 Å and of the long side from 4.47 to 4.67 Å. The corresponding Nb rectangle in NbC does not have a Nb atom in the center as compared to the Nb rectangle shown in Fig. 1(e), however, a view of the interface from a complementary (110) direction provides a close match to the TEM image in Fig. 1(a) because the out-of-plane Nb atoms can also be seen in the center of the rectangle as shown in Fig. 1(e). Since the in-plane deformation for this model is small, the focus of the analysis of this model will be on the difference between the phonon DOS for atoms at an extended, commensurate interface vs. bulk Nb and bulk NbC.

First, consider the phonon dispersions in bulk NbC. Phonon dispersion curves for NbC in a B1 NaCl structure were generated by density functional theory calculations performed within the Quantum Espresso package [29] using plane waves and ultrasoft pseudopotentials [30] along with density functional perturbation theory [31] and harmonic approximation for the phonon calculations. Exchange-correlation effects were treated using the generalized gradient approximation according to Perdew-Burke-Ernzerhof [32]. The k-mesh used was a 24x24x24 Monkhorst-Pack grid and the q-mesh for the phonons was an 8x8x8 grid. Occupation numbers were determined using the Methfessel-Paxton scheme with a broadening parameter σ=0.025 Ry and a plane wave energy cutoff of 40 Ry was also used.

Comparison between the calculated dispersions of the bulk NbC modeled with a 4.47 Å lattice parameter and the two-phonon Raman signal show modest overall agreement; however, better agreement with the two-phonon Raman signal is found for the model of NbC compressed to 88% of its ideal volume, since this compression shifts the frequencies of the acoustic phonons up by ~2 THz. As discussed below, the acoustic phonon DOS generated by DFT for this uniform compression is also in good agreement (including the shape) with the measured first order Raman spectrum in Figure 3(b). The calculated acoustic phonon dispersion curves for the compressed NbC model and the two-phonon Raman signal are shown in Fig. 4. Each peak of the two-phonon Raman (with frequency axis divided by 2) corresponds to regions of phonon anomalies. In particular the doublet near 7 THz appears to be due to the close frequencies of the strong phonon dips along the Γ-X and Γ-K directions as well as the L point in the Brillouin zone. The two lower peaks in the range 5-6 THz also correlate well with regions of phonon dips. A small peak near 8 THz tentatively corresponds to the dip at the W point but it is also possible that it corresponds to an A+A' overtone. In effect, all of the regions where strong phonon renormalization is occurring correspond to a two-phonon Raman peak. To our knowledge, such a correlation has not been achieved before on any superconducting transition metal compound.

On the other hand, we approximated a strong, anisotropic deformation, which potentially breaks symmetry and might result in additional Raman peaks, by a uniform, bulk, isotropic compression. Also, such a compression should result in an increase in optical as well as the acoustic phonon frequencies, which can be seen in the DOS plots (solid black curves vs. broken red curves) presented in the bottom panel of Fig. 3(b). The first order Raman spectra do not reveal the expected shift of the optical modes despite the excellent agreement between measured Raman and calculated acoustic mode DOS. This is puzzling and therefore we have also calculated the phonon DOS for the second structural model, the Nb(110)/NbC(110) interface.

We modeled the interface structure using a six layer thick slab of bcc Nb cut along the (110) plane matched to a four layer thick slab of fcc NbC cut along the (110) plane as shown in Fig. 1 (e), with periodic boundary conditions. Density functional theory as implemented in VASP [33] was used to optimize the geometry and density functional perturbation theory was used to calculate the force constants. Phonopy [34] was used to produce the phonon DOS. The peak widths and positions of the interface model for both the acoustic and optical one-phonon regions match well with the observed Raman spectra, as shown in the bottom panel of Fig. 3(c). Both optical and acoustic regions are broadened for the interface model when compared to bulk Nb and NbC. Partial DOS in Fig. 3(c) indicate that the phonon states associated with Nb close to the interface are responsible for broadening of the acoustic one-phonon peaks. An analysis of optimized geometries indicates the presence of shorter Nb-C distances at the first interface layer, which correlates with the acoustic phonon states observed at higher frequencies. The agreement between the upshifted acoustic frequencies in this model and the Raman signal for the one-phonon frequencies should also be reflected in the location of the two-phonon signal. Defects in NbC were proposed as a possible explanation of narrower and more centrally peaked two-phonon spectra relative to one-phonon spectra [13]. At the same time, vacancies were shown not to enhance the two-phonon peaks, based on comparing spectra for different carbon deficiencies. However, there is a possibility that interfacial structures of NbC precipitates could lead to enhancement of two-phonon Raman signal, which correlates well with our measurements in Fig. 3.

To summarize this section, the upward shift of the acoustic modes in the first order Raman compared to the phonon DOS measured by neutron scattering, suggest that the NbC inclusions may be under compressive strain. Choosing a volume V = 0.88 of bulk NbC (uniform compression) leads to the theoretical shift of the acoustic mode DOS in good agreement with experiment and at the same time provides phonon dispersions which demonstrate a correlation between regions of phonon anomalies and two-phonon Raman peaks. While such agreement is compelling, there is no observed strong shift of the Raman optical modes as would be expected in this model. A second model that matches Nb (110) and NbC (110) leads to a relatively smaller isotropic tensile strain. Modeling the extended interface using DFT leads to a broadening and upward shift of the acoustic mode DOS of NbC, but has relatively minor effect on the optical mode DOS, in agreement with first order Raman. In this model the sharp two-phonon signal might originate in the local vibrational modes in particular planes of the extended interface between Nb and NbC.

## V. EVIDENCE OF AN INCREASED SUPERCONDUCTING GAP FROM POINT CONTACT TUNNELING

The origin of the strong two-phonon Raman in these NbC precipitates is not fully understood at present. One possibility is that this is the intrinsic response of NbC and bulk samples simply haven't revealed it, instead showing reduced, broadened features due possibly to surface imperfections. The bulk NbC Raman spectra were variable and sensitive to surface preparation and polishing methods [13]. The other possibility is that these precipitates are showing a real enhancement of the two-phonon Raman, above that of bulk NbC. Given that the sizes of the precipitates are variable, it does not seem to be a simple size effect causing the enhancement. It might originate in the strain (either compressive or tensile) that these fcc NbC inclusions are under as a result of their coherency with the host Nb matrix or in the local structures at the interfaces between Nb and NbC inclusions. Considering theoretical models [14] the two phonon strength is tied to the electron-phonon spectral function [16], $\alpha^2 F(\omega)$, and there is the possibility that this spectral function may have also been enhanced in these NbC inclusions. Hence, they may even have a larger $T_c$ than bulk NbC.

To test this possibility, tunneling measurements using a mechanical contact were initiated. Patches exhibiting NbC Raman spectra were found frequently on the sample A. A region rich in such patches, approximately 20 by 50 micron in size, was identified under Raman microscope and its location marked on the sides of the sample. We then aligned a PtIr tip on top of the region of interest and searched for tunnel junctions through a point contact method [35]. Near stoichiometric NbC should have a superconducting $T_c$ close to 12 K [36]. Using this value and assuming a similar strong coupling ratio, $2\Delta/kT_c = 3.9$, as that of Nb (with $\Delta = 1.55$ meV), an estimated $\Delta = 2.0$ meV for the expected gap parameter of NbC is obtained. Some tunneling spectra with large energy gaps were observed as shown in Fig. 5. In general, these large gap spectra are relatively broad and exhibit a double gap feature with the small gap resembling that observed for a pure Nb region outside the patch, (bottom red curve of Fig. 5). The multigap feature could be interpreted as the tip covering both a NbC patch and the surrounding Nb, forming parallel junctions. However, as mentioned earlier, TEM studies show that the top surface above a NbC inclusion often does not exhibit a Nb oxide layer, (e.g. Fig. 1(a)), and thus would not be expected to have a native, insulating tunnel barrier. Thus the two-gap feature may be the result of some type of proximity effect. Despite this uncertainty, single gap fits of the larger gap feature were performed and these lead to $\Delta$ values in the range of 2.1 meV to 3.3 meV among the three junctions in Fig. 5. The value at the low end would be consistent with estimates for bulk NbC, but those at the high end would suggest a local NbC phase with $T_c$ significantly higher than 12 K and approaching 20 K. Some junctions (not shown) were formed and measured near the $T_c$ of Nb and while the smaller Nb gap feature disappeared the larger gap was still visible.

## VI. CONCLUSIONS

In summary, high purity Nb, subjected to one or more of the processing steps of strain, etching and annealing used in the fabrication of SRF cavities, consistently reveals micron sized surface regions that contain excess carbon. The new results of this study show that, in addition to amorphous C and chain-type hydrocarbons observed earlier by Raman, these patches often reveal spectra consistent with bulk NbC but with a significantly enhanced two-phonon Raman signal. TEM/EELS measurements reveal nanoscale precipitates of NbC at the surface of processed Nb, corroborating the Raman results. High resolution TEM reveals no observable, significant change in lattice constant of the Nb indicating the NbC is coherent with the Nb host. Considering the strong possibility that the coherency is linked to the enhanced two-phonon Raman we have examined in more detail the consequences to the phonon modes. Two possible lattice matchings of high symmetry, close packing planes have been considered, one leading to anisotropic compressive strain Nb(110)/NbC(111), and the other isotropic, weaker tensile strain Nb(110)/NbC(110). Either type of lattice matching seems plausible and is not ruled out by the atomically resolved TEM. A direct comparison to DFT calculations of the NbC phonon dispersions (under uniform compressive strain), shows directly that peaks in two-phonon Raman correspond to regions of phonon softening due to e-ph renormalization. Thus, theoretical notions that the two-phonon Raman strength is tied to phonon anomalies and the electron-phonon spectral function, $\alpha^2F(\omega)$, have been substantiated. But this model shows discrepancies with the optical mode DOS observed in first order Raman. An extended interface model for Nb(110)/NbC(110) is in good agreement with the first order Raman, and also suggests the presence of local vibrational modes at the interface that can give rise to sharp two-phonon Raman features. This interface model explains the upward frequency shift of the acoustic modes and the corresponding insensitivity of the optic modes as is observed. At the present time, there is not enough information to pick one model over the other. In either case the strong two-phonon Raman is a consequence of strong electron-phonon interaction and there is the possibility that $\alpha^2F(\omega)$ may have been enhanced in these NbC inclusions, a result supported by preliminary tunneling data. If so, then this suggests that there may be new pathways for manipulating the e-ph interaction, and raising $T_c$, in transition metal carbides.

## ACKNOWLEDGEMENTS

The authors thank G. Ciovati of Jefferson Laboratory for supplying Nb samples used in this study. Calculations (H.L.) were performed with financial support by the SSF-project Designed multicomponent coatings, MultiFilms and the Swedish Research Council. Calculations were carried out at the Swedish National Infrastructure for Computing (SNIC), Argonne LCRC and Argonne Center for Nanoscale Materials. The work at Argonne National Laboratory and the use of the Center for Nanoscale Materials and the Electron Microscopy center at Argonne National Laboratory were supported by the U.S. Department of Energy, Office of Science, Office of Basic Energy Sciences under contract No. DE-AC02-06CH11357. This work was also supported by the Department of Energy, Office of Science, Office of High Energy Physics, early career award FWP #50335 to T.P.

**References**

[1]  L. E. Toth, *Transition Metal Carbides and Nitrides* (Academic Press, New York, 1971).

[2]  J. Zmuidzinas, Annu. Rev. Condens. Matter Phys. **3**, 169 (2012).

[3]  J. A. Klug, N. G. Becker, N. R. Groll, C. Cao, M. S. Weimer, M. J. Pellin, J. F. Zasadzinski, and T. Proslier, Appl. Phys. Lett. **103**, 211602 (2013).

[4]  R. Romestain, B. Delaet, P. Renaud-Goud, I. Wang, C. Jorel, J.-C. Villegier, and J.-P. Poizat, New Journal of Physics **6**, 129 (2004).

[5]  Z. Wang, D. Liu, S.-L. Li, J. Li, and S.-C. Shi, Supercond. Sci. Technol. **27**, 075003 (2014).

[6]  C. M. Varma and W. Weber, Phys. Rev. B **19**, 6142 (1979).

[7]  B. M. Klein, L. L. Boyer, and D. A. Papaconstantopoulos, Solid State Commun. **20**, 937 (1976).

[8]  D. Reznik, Advances in Condensed Matter Physics **2010,** 523549 (2010).

[9]  S. Blackburn, M. Cote, S. Louie, and M. L. Cohen, Phys. Rev. B **84**, 104506 (2011).

[10] E. I. Isaev, S. I. Simak, I. A. Abrikosov, R. Ahuja, Y. K. Vekilov, M. I. Katsnelson, A. I. Lichtenstein, and B. Johansson, J. Appl. Phys. **101**, 123519 (2007).

[11] E. G. Maksimov, M. V. Magnitskaya, S. V. Ebert, and S. Yu. Savrasov, JETP Lett. **80,** 548 (2004).

[12] Z. P. Yin, A. Kutepov, and G. Kotliar, Phys. Rev. X **3**, 021011 (2013).

[13] H. Wipf, M. V. Klein, and W. S. Williams, Phys. Stat. Sol. (b) **108**, 489 (1981).

[14] M. V. Klein, Phys. Rev. B **24**, 4208 (1981).

[15] M. V. Klein and S. B. Dierker, Phys. Rev. B **29**, 4976 (1984).

[16] See for example, E.L. Wolf, *Principles of Electron Tunneling Spectroscopy,* second edition, (Oxford press, 2011). Whereas tunneling spectroscopy can measure $\alpha^2 F(\omega)$ the two-phonon Raman is roughly proportional to $\alpha^4 F(\omega)$, Miles Klein (private communication).

[17] R. Kaiser, W. Spengler, S. Schicktanz, and C. Politis, Phys. Stat. Sol. (b) **87**, 565 (1978).

[18] C. Cao, et al., Phys. Rev. ST Accel. Beams **16,** 064701 (2013).

[19] A. Romanenko, G. Wu, and L. D. Cooley, in *Proceedings of SRF2011, Chicago* (2011) THPO008.

[20] A. Dzyuba and L. D. Cooley, Supercond. Sci. Technol. **27**, 035001 (2014).

[21] P. Maheshwari, F. A. Stevie, G. Myeneni, G. Ciovati, J. M. Rigsbee, and D. P. Griffis, in *First International Symposium on the Superconducting Science and Technology of Ingot Niobium, Newport News* (AIP Publishing, 2011), Vol. 1352, p. 151.

[22] Z. C. Szkopiak and A. P. Miodownik, J. Nucl. Mater. **17**, 20 (1965).

[23] T. R. Bieler, et al., Phys. Rev. ST Accel. Beams **13**, 031002 (2010).

[24] L. D. Cooley, et al., IEEE T. Appl. Supercon. **21**, 2609 (2010).

[25]   M. Uz and R. H. Titran, AIP Conf. Proc. **217**, 172 (1991).

[26]   X. Zhao, G. Ciovati, and T. R. Bieler, Phys. Rev. ST Accel. Beams **13**, 124701 (2010).

[27]   W. Spengler and R. Kaiser, Solid State Commun. **18**, 881 (1976).

[28]   G. K. Tirumalasetty, et al., Acta Mater. **59**, 7406 (2011).

[29]   P. Giannozzi, et al., J. Phys. Condens. Matter **21**, 395502 (2009).

[30]   D. Vanderbilt, Phys. Rev. B **41**, R7892 (1990).

[31]   S. Baroni, S. D. Gironcoli, A. D. Corso, and P. Giannozzi, Rev. Mod. Phys. **73**, 515 (2001).

[32]   J. P. Perdew, K. Burke, and M. Ernzerhof, Phys. Rev. Lett. **77**, 3865 (1996).

[33]   G. Kresse and J. Furthmuller, Phys. Rev. B **54**, 11169 (1996).

[34]   A. Togo, F. Oba, and I. Tanaka, Phys. Rev. B **78**, 134106 (2008).

[35]   L. Ozyuzer, J. F. Zasadzinski, and K. E. Gray, Cryogenics **38**, 911 (1998).

[36]   R. Jha and V. P. S. Awana, J. Sup. Nov. Mag. **25**, 1421 (2012).

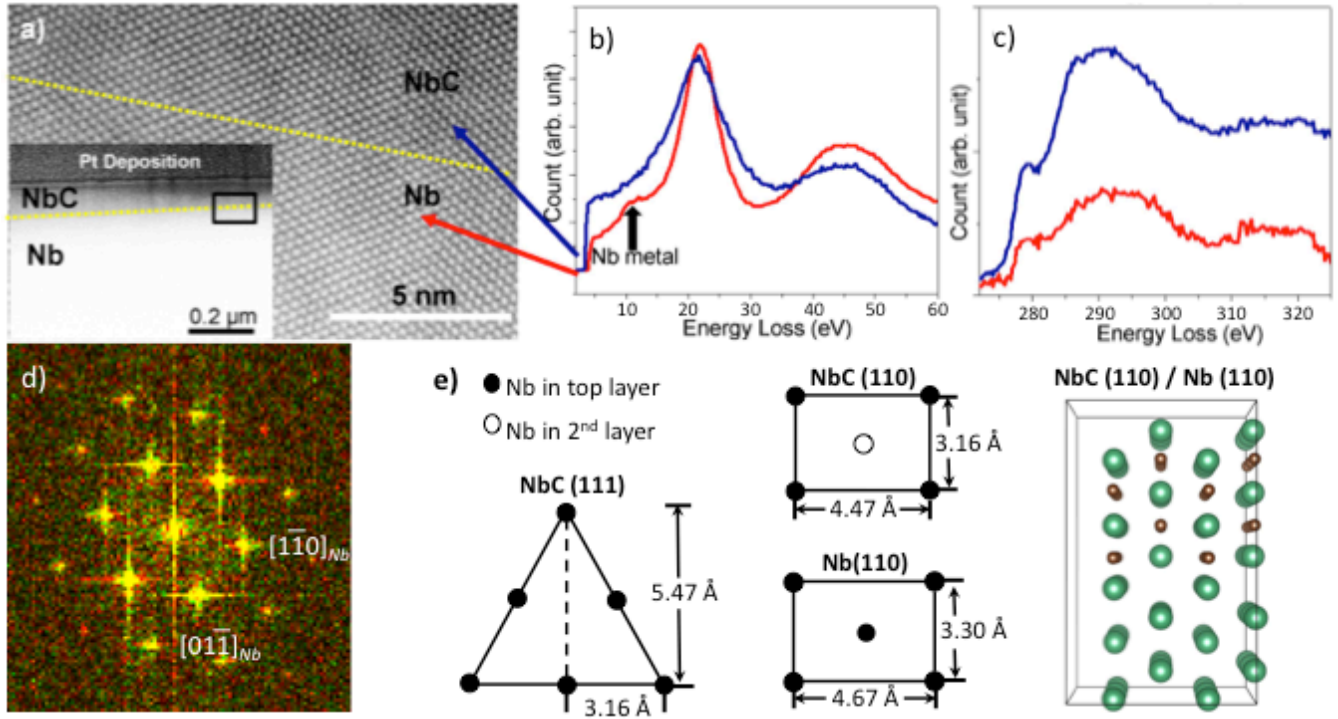

FIG. 1. (a) Inset shows an aberration-corrected STEM annular dark field image of surface of Nb rod sample #2 which shows ~ 90 nm thick region with a different contrast compared to the bulk Nb. Main panel shows atomic-resolution image of the interface between the surface inclusion and the Nb bulk, indicated by the broken line. The atomic plane of bcc Nb is observed with no evidence of any change in lattice constant across the boundary. (b) Low-loss EELS spectrum of Nb (red curve) and NbC (blue curve) regions. A shoulder at ~10 eV, characteristic of Nb is absent in the region near the surface which is likely NbC. (c) Core-loss EELS spectra for Nb (red curve) and NbC (blue curve). (d) Fourier transforms (FFT) of the Nb and NbC regions, where the FFT of Nb is shown in green and the NbC in red. The overlap is indicated by yellow. (e) Nb atom arrangements for the (111) and (110) planes of NbC, the (110) plane of Nb, and an atomic model of the interface between NbC (110) / Nb (110). In the atomic model, Nb atoms are represented as large green spheres and C atoms are small brown spheres.

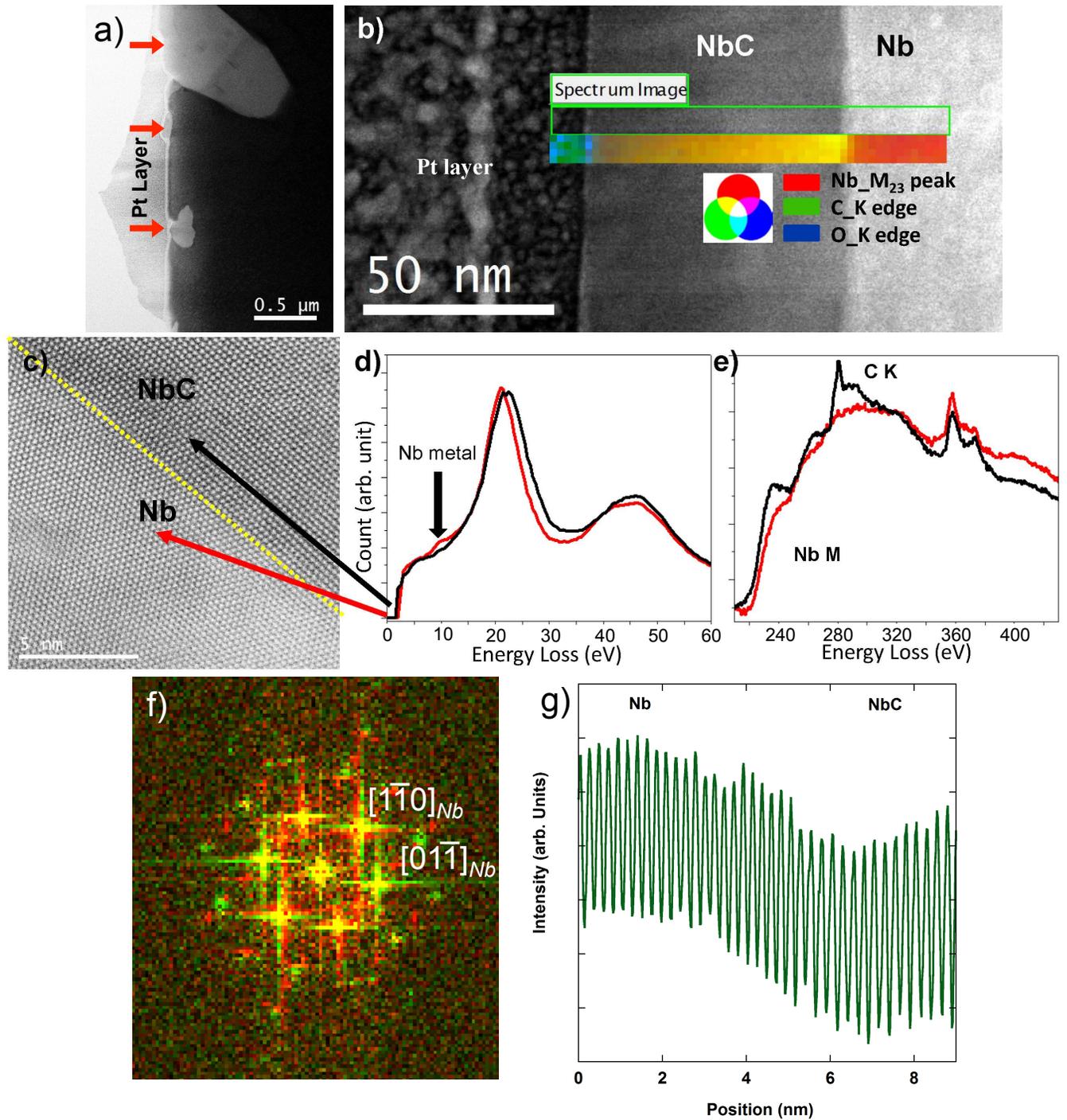

FIG. 2. (a) Annular bright field (ABF) image from sample #1. Three NbC inclusions (indicated by arrows) are observed varying in linear dimension from ~1 μm long (top) to ~60 nm across (middle). (b) ABF image of the middle inclusion with composition of the Nb bulk and the inclusion indicated by the integrated EEL spectrum intensities. (c) Atomic-resolution image of the interface between the bottom inclusion and the Nb bulk. (d-e) Low-loss and core-los EEL spectra from the two regions indicated in (c). The black EELS curve shows peaks consistent with NbC including the sharp C peak centered at 288 eV. (f) Fourier transforms (FFT) of the Nb and NbC regions, where the FFT of Nb is shown in green and the NbC in red. The overlap is indicated by yellow. (g) Image intensity profile across the hetero-interface, perpendicular to the Nb(110) planes.

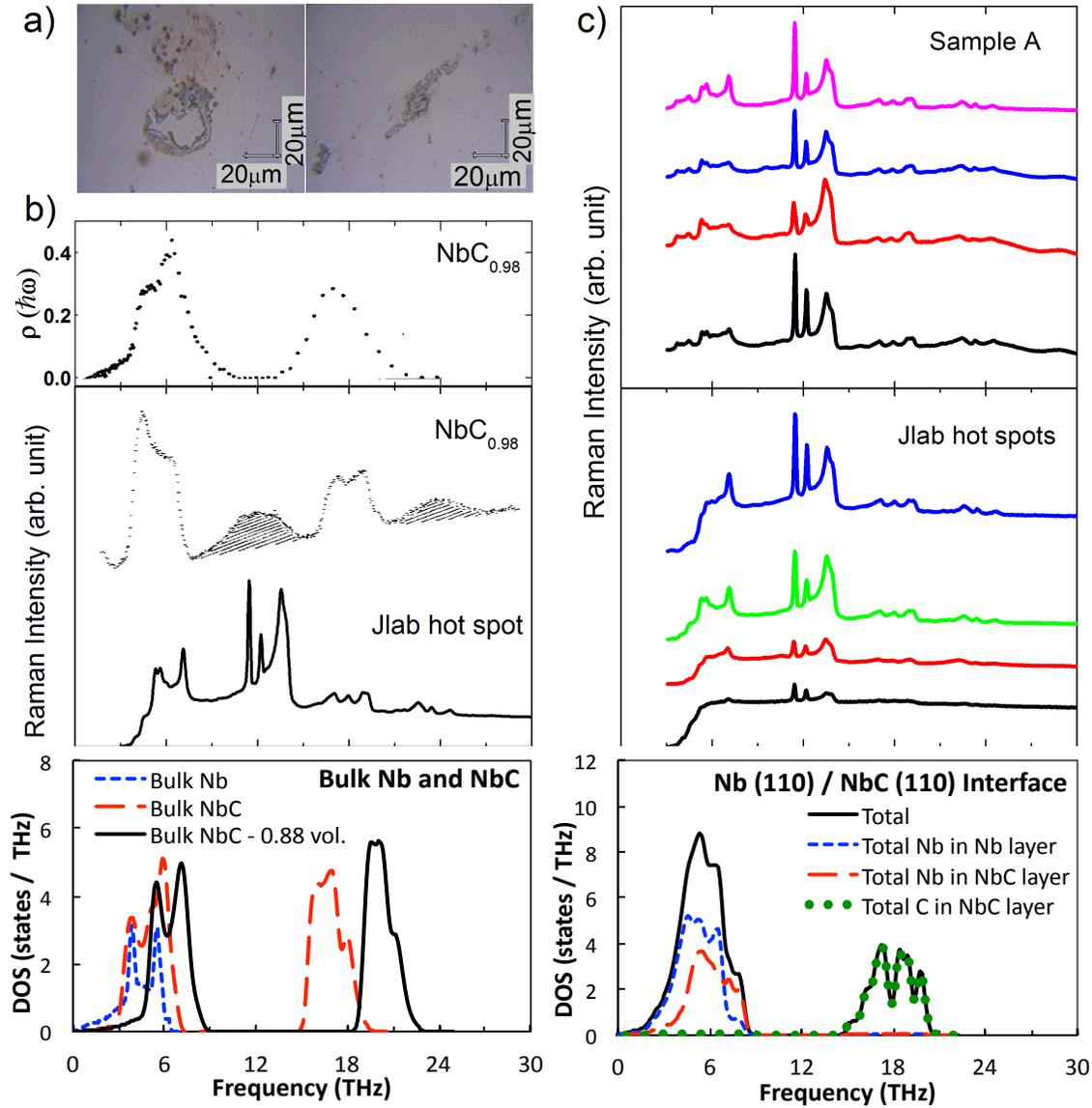

FIG. 3. (a) Microscope camera photos of surface patches on JLab sample A. (b) A representative Raman backscatter spectra shown in comparison to Raman data (ref. 13), the phonon DOS of NbC$_{0.98}$ measured from neutron scattering (ref. 27) and the phonon DOS of bulk Nb, NbC, and compressed NbC calculated using DFT. (c) Representative Raman spectra from JLab sample A, cavity pieces cut out from hot spots, and total and partial phonon DOS for the Nb(110)/NbC(110) interface model calculated using DFT. Dashed shaded regions in (b)

correspond to two-phonon A+A near 12 THz and A+O regions near 24 THz. Raman spectra in (c) are highly reproducible and characterized by four peaks centered near 12 THz.

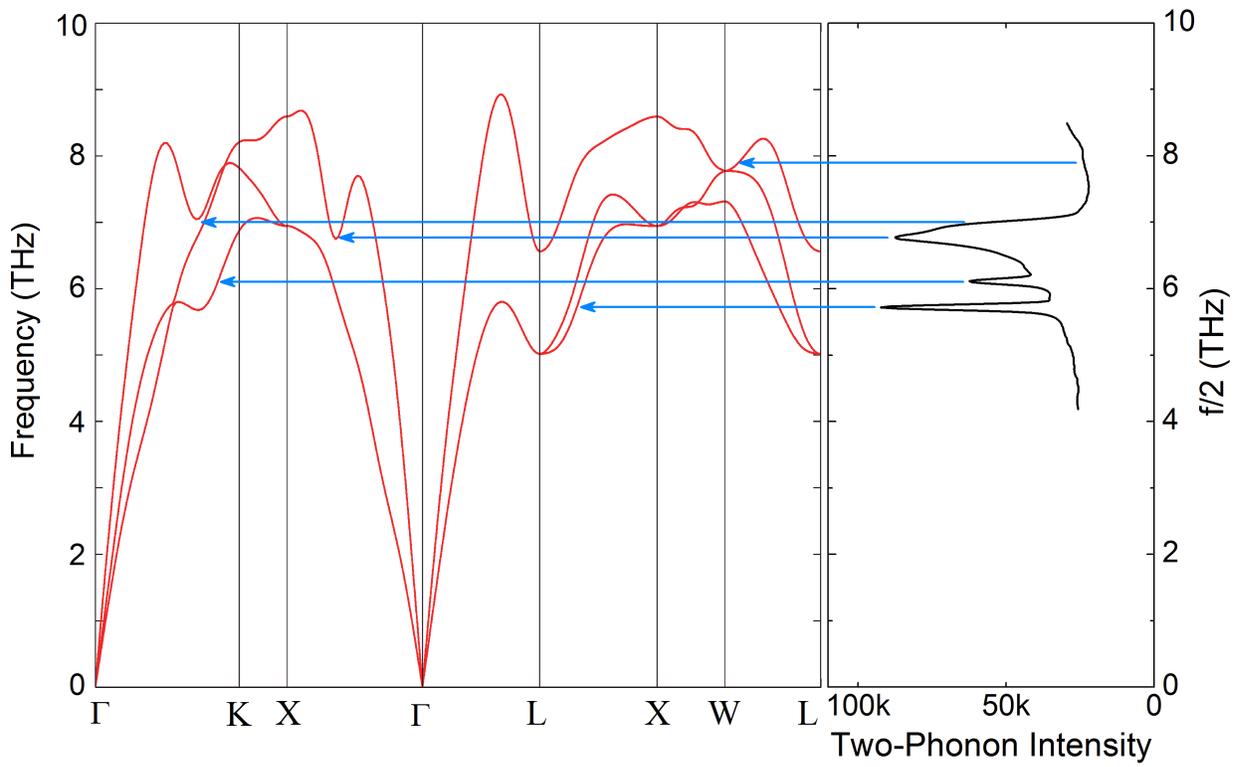

FIG. 4. Comparison of calculated phonon dispersions for NbC under uniform volume compression of 0.88 of the bulk volume to a representative two-phonon Raman spectrum from a processed Nb sample. Arrows suggest the regions of strong phonon renormalization responsible for the two-phonon peaks.

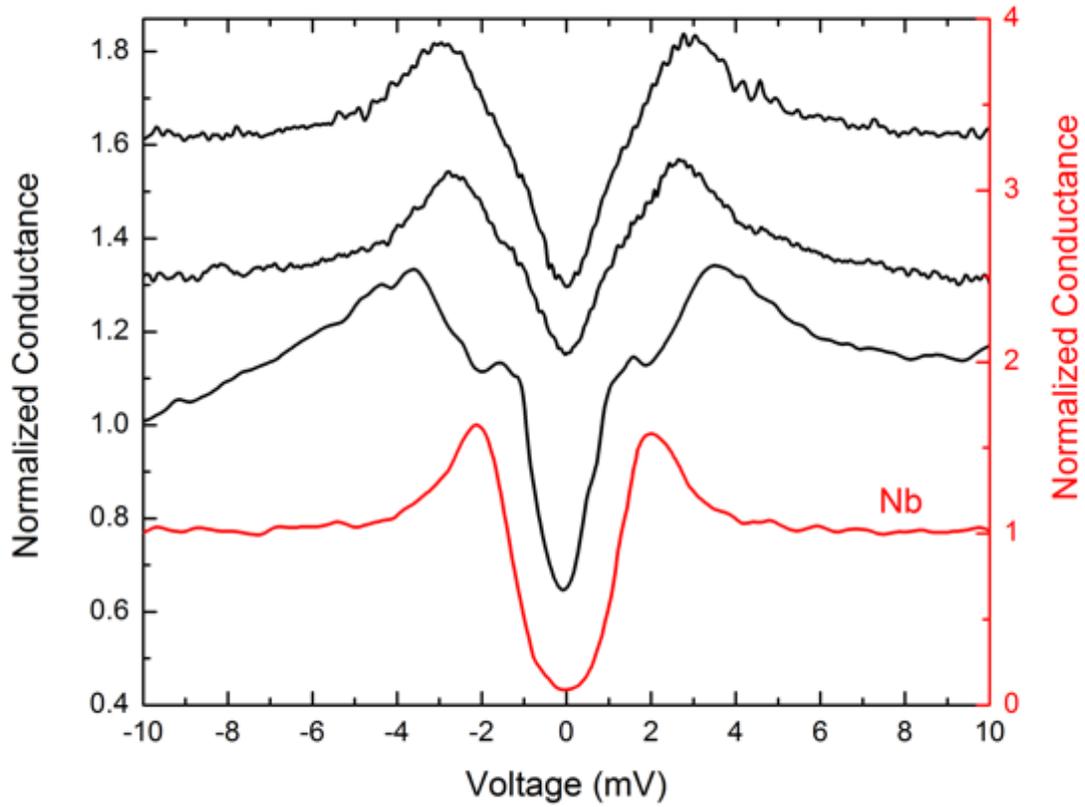

FIG. 5. Normalized tunneling conductances from three large-gap PCT tunnel junctions (black curves) obtained near a surface patch that typically displays NbC in the Raman spectrum. Regions ouside the patch display tunneling conductances typical of pure Nb (red curve) which is close to the sub-gap feature seen in the black curves.